\documentclass[utf8]{FrontiersinHarvard} 

\usepackage{url,hyperref,lineno,microtype,subcaption}
\usepackage[onehalfspacing]{setspace}

\long\def\comment#1{}

\newcommand{\ltsim}{\raisebox{-1.0ex}{$\stackrel{\textstyle<}{\sim}$}}
\long\def\comment#1{}
\def\kms{km~s$^{-1}$}
\def\al{Alfv\'{e}n}

\def\goes{{\sl GOES}}
\def\yohkoh{{\sl Yohkoh}}

\def\hinode{{\sl Hinode}}
\def\sdo{{\sl SDO}}
\def\stereo{{\sl STEREO}}
\def\iris{{\sl IRIS}}
\def\goes{{\sl GOES}}

\def\trace{{\sl TRACE}}
\def\so{{\sl Solar Orbiter}}
\def\psp{{\sl Parker Solar Probe}}

\def\fex{Fe~{\sc x}}
\def\feix{Fe~{\sc ix}}

\def\caii{Ca~{\sc ii}}

\def\keyFont{\fontsize{8}{11}\helveticabold }
\def\firstAuthorLast{Sterling {et~al.}} 
\def\Authors{Alphonse C. Sterling\,$^{1,*}$, Ronald L. Moore\,$^{2,1}$, Navdeep K. Panesar$^{3,4}$, Tanmoy Samanta$^{5}$,
Sanjiv K. Tiwari$^{3,4}$ and Sabrina L. Savage\,$^{1}$}
\def\Address{$^{1}$NASA, Marshall Space Flight Center, Huntsville, AL 35812, USA \\
$^{2}$Center for Space Plasma and Aeronomic Research, University of Alabama in Huntsville, Huntsville, AL 35805, USA \\
$^{3}$Lockheed Martin Solar and Astrophysics Laboratory, 3251 Hanover Street Building 252, Palo Alto, CA 94304, USA \\
$^{4}$Bay Area Environmental Research Institute, NASA Research Park, Moffett Field, CA 94035, USA \\
$^{5}$Indian Institute of Astrophysics, Koramangala, Bangalore 560034, India
}
\def\corrAuthor{A. C. Sterling}

\def\corrEmail{alphonse.sterling@nasa.gov}

\begin{document}
\onecolumn
\firstpage{1}

\title[Small-Scale Explosive Solar Features]{Future High-Resolution and High-Cadence Observations for Unraveling Small-Scale Explosive Solar Features} 

\author[\firstAuthorLast ]{\Authors} 
\address{} 
\correspondance{} 

\extraAuth{}

\maketitle

\begin{abstract}

\section{}
Solar coronal jets are frequently occurring collimated ejections of solar plasma, originating from magnetically mixed polarity 
locations on the Sun of size scale comparable to that of a supergranule.  Many, if not most,  coronal jets are produced by eruptions
of small-scale filaments, or minifilaments, whose magnetic field reconnects both with itself and  also with surrounding coronal 
field.  There is evidence that minifilament eruptions are a scaled-down version of typical filament eruptions that produce solar flares 
and coronal mass ejections (CMEs). Moreover, the magnetic processes building up to and triggering minifilament 
eruptions, which is often flux cancelation, might similarly build up and trigger the larger filaments to erupt.  Thus, detailed study of coronal jets 
will inform us of the physics leading to, triggering, and driving the larger
eruptions.  Additionally, such studies potentially can inform us of smaller-scale coronal-jet-like features, such as jetlets and perhaps some 
spicules, that might work the same way as coronal jets.  We propose a high-resolution ($\sim$0$''\kern-0.5em.\,1$ pixels), 
high-cadence 
($\sim$5\,seconds) EUV-solar-imaging mission for the upcoming decades, that would be dedicated to 
observations of features of the coronal-jet size scale, and smaller-scale solar features produced by similar physics.  Such a mission could provide invaluable insight into the operation
of larger features such as CMEs that produce significant Space Weather disturbances, and also smaller-scale features that could
be important for coronal heating, solar wind acceleration, and heliospheric features such as the  magnetic switchbacks that are 
frequently observed in the solar wind.

\tiny
 \keyFont{ \section{Keywords:} Solar filament eruptions, solar corona, solar x-ray emission, solar extreme ultraviolet emission, solar coronal jets, solar magnetic activity} 
\end{abstract}

\section{Introduction}
\label{sec-introduction}

Solar coronal jets are transient phenomena that originate near the solar surface and extend
into the corona in the form of long and narrow spires.  They are visible at soft X-ray (SXR) and EUV wavelengths, 
and occur in coronal holes, 
quiet Sun, and the periphery of active regions.  While there were some scattered earlier observations, coronal jets were
first observed in detail in SXRs with the \yohkoh\ satellite \citep{shibata.et92}, and they have since been observed
with several instruments in different wavelengths.  (Surges, which traditionally have been observed in the chromosphere, share some properties with coronal jets,
and in some cases accompany coronal jets; e.g., \citeauthor{canfield.et96}~\citeyear{canfield.et96}; \citeauthor{moore.et10}~\citeyear{moore.et10},~\citeyear{moore.et13}; \citeauthor{sterling.et16b}~\citeyear{sterling.et16b}.)
In coronal images, coronal jets consist of a spire that emanates from a bright base region and extends into the 
corona.  From a study of SXR coronal jets in polar coronal holes \citep{savcheva.et07}, coronal jets
live for tens of minutes, and have an 
occurrence rate of about 60/day in the two polar coronal holes, which translates to about one per hour in a
given polar coronal hole.  That same study reported that the coronal-jet spires reach $\sim$50,000\,km, and have 
widths $\sim$10,000\,km.  Coronal jets seem to have two speeds, or two components of different speeds; a slower speed 
of $\sim$200\,\kms, which is close to the local sound speed, and a faster speed of $\sim$1000\,\kms, which is near
the \al\ speed \citep{cirtain.et07}.  It had been pointed out much earlier \citep{shibata.et92} that the base region of the coronal jet
often has a particularly bright spot in the coronal-jet base, and that brightening is offset to one side of the base.  
Several general summaries and reviews of coronal jets have come out during different
eras of coronal-jet studies \citep[][]{shimojo.et00,shibata.et11,raouafi.et16,hinode.et19,shen21,sterling21,schmieder22}.

It has been argued by \citet{sterling.et15} that most or all coronal jets result from the eruption of small-scale 
filaments, or {\it  minifilaments} (being
several times to orders-of-magnitude smaller than ``typical" large-scale filaments, that erupt to make typical solar flares and
coronal mass ejections), based on their study of 20 polar coronal hole jets, and building on several earlier studies of
coronal jets \citep{nistico.et09,moore.et10,moore.et13,hong.et11,huang.et12,shen.et12,adams.et14}. \citet{sterling.et15} 
proposed that the minifilament eruption and resulting coronal jet are scaled-down versions of typical filament eruptions
that produce solar flares and coronal mass ejections (CMEs). 

\begin{figure}[h!]
\begin{center}
\includegraphics[angle=0,width=12cm]{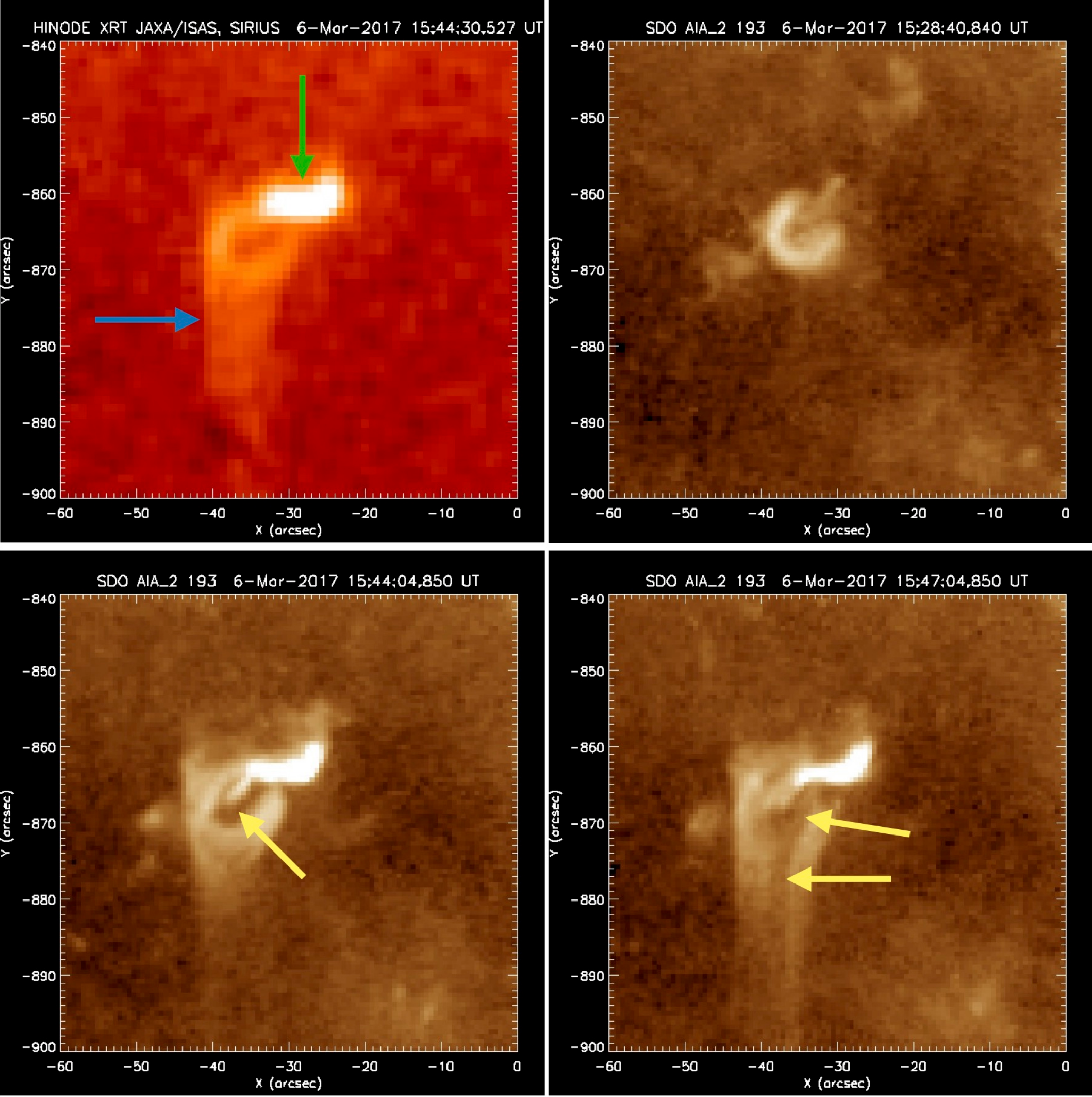}
\end{center}
\caption{Example of a coronal jet, occurring on 2017 March 6.   (a) \hinode/XRT soft X-ray image of the coronal-jet spire (blue arrow) and
the jet bright point (``JBP") brightening in the coronal-jet's base (green arrow).  North is upward and west is to the right in this and all 
solar images in this paper.  This coronal jet is in the south polar region, and hence the spire is pointing downward.}\label{fig:s22_jet14}
\end{figure}

Figure~\ref{fig:s22_jet14} shows an example of a coronal jet, occurring in the south polar coronal hole.  
Panel~\ref{fig:s22_jet14} (a) shows the coronal jet 
when it is well developed, in a SXR image from the \hinode\ spacecraft.  The blue arrow points to the spire, and the 
green arrow points to the base brightening.  We call the latter a JBP for the ``jet bright point" in the coronal-jet's base 
\citep[this terminology is from][]{sterling.et15}.  Panels~\ref{fig:s22_jet14} (b)---\ref{fig:s22_jet14} (d) show the 
same location in EUV images 
from the \sdo/AIA~instrument's 193\,\AA\ channel.  Panel~\ref{fig:s22_jet14}(b) shows the situation prior to the start of the coronal jet.  
Panel~\ref{fig:s22_jet14} (c) is from about the same time as the SXR image in~\ref{fig:s22_jet14} (a).  It shows the spire less prominently than in 
SXRs at that moment, but it also shows a minifilament (in absorption) in the process of erupting outward (yellow arrow).  These erupting 
minifilaments usually are, at best, only hinted at in SXR images, but are often clearly visible in at least one and often in 
several AIA EUV channels \citep{sterling.et22}.
In~\ref{fig:s22_jet14} (d), this minifilament has continued to erupt, with portions of it leaking out into the bright spire.

\begin{figure}[h!]
\begin{center}
\includegraphics[angle=90,width=14cm]{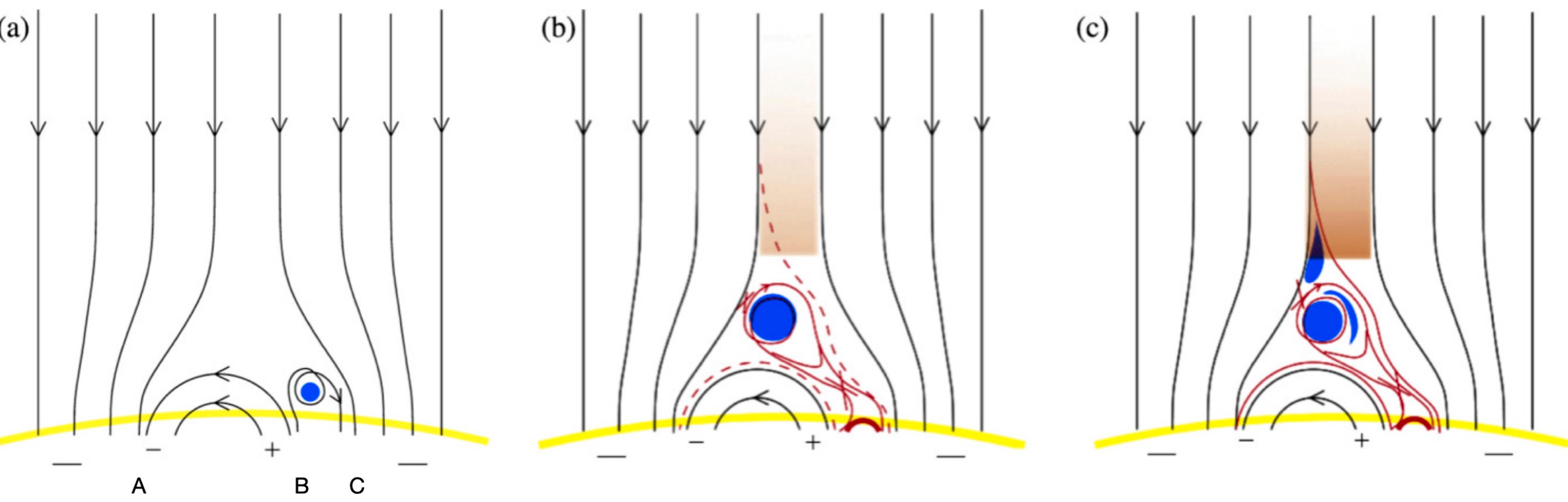}
\end{center}
\caption{Schematic showing the minifilament-eruption mechanism for coronal jets, from \citet{sterling.et15} \citep[this slightly
modified version of the original figure appeared in][]{sterling.et18}. 
This is a two-dimensional cross-sectional cut of the jetting region, with the yellow curve representing 
the Sun's limb.  Black lines represent magnetic field lines prior to reconnection, 
with the arrow heads and the ``+" and ``-" signs indicating polarities.  Red lines indicate reconnected field
lines, with the dotted red lines indicating that the field lines are newly reconnected.  The blue circle 
indicates cool minifilament material, with the circular field surrounding it  being the minifilament's 
field. Panel (a) shows the situation prior to eruption.  ``A," ``B," and ``C" indicate locations 
for reference.  In (b) the minifiament's field is erupting, carrying with it the cool minifilament material.
Red X-es show reconnections, with the one below the erupting minifilament being flare-like reconnection \citep[called
``internal" reconnection, being internal to the erupting flux-tube's lobe field;][]{sterling.et15}
resulting in the jet bright point (JBP), represented by the thick-red simicircle between the 
B and C locations.  The reconnection in front of (to the left and slightly above the center of) the 
erupting minifilament (``external" reconnection) results in new open field,
along which heated plasma flows to form the coronal jet's spire (shaded region). This reconnection also adds
new heated closed field above the magnetic lobe between locations A and B\@.   In (c) the higher-height
reconnection has proceeded enough that some of the cool minifilament material is flowing out 
onto the spire on newly reconnected open field.}\label{fig:s15_zu}
\end{figure}

Figure~\ref{fig:s15_zu} shows the scenario proposed by \citet{sterling.et15} to explain coronal jets.  This 
shows a coronal-hole region: most of the photospheric magnetic flux has the same polarity (negative in this case), and the 
ambient coronal field is open. 
While this drawing is tailored for a coronal hole region, the same description holds where 
the ambient coronal field is a long loop (compared to the size of the base region) instead, which would be common in 
quiet Sun and active regions.  A positive-polarity patch is present in the region, and this forms an anemone-type structure \citep{shibata.et07}, with flux emanating out of the positive patch and closing down in negative flux surrounding the 
positive patch in 3-D, where Figure~\ref{fig:s15_zu}(a) shows a 2-D cross-section of this structure. One 
lobe of the anemone - 
the smaller lobe on the right-hand side (i.e., the closed-field region between B and C in Fig.~\ref{fig:s15_zu}) in 
this depiction - contains non-potential (sheared and twisted) field, and holds a 
minifilament.  The adjacent lobe on the left
side is larger and contains more-nearly potential field.  Figure~\ref{fig:s15_zu}(b) shows the minifiament, and its enveloping
flux-rope field, erupting.  This results in two magnetic reconnections.  One reconnection occurs where 
the erupting-minifilament's field encounters the ambient coronal field above the larger lobe.  This adds new, heated loops 
to the large lobe, and creates new open field, along which heated material can flow out and create the coronal-jet spire.  The 
second reconnection occurs among the leg field of the erupting flux-rope-enveloping field.  This causes a miniature flare to occur, 
in a fashion analogous to how typical solar flares are formed \citep[e.g.,][]{hirayama74,shibata.et95,moore.et01}. This miniature flare
is what appears as the aforementioned JBP.\@  In Figure~\ref{fig:s15_zu}(c), the erupting field has reconnected far enough into the open-field region for the 
cool minifilament material in the flux-rope core of the erupting field to leak out onto the open field, where it flows away as part of the spire.

Studies of coronal jets occurring on the solar disk have provided insight into the magnetic origins of the coronal-jet-producing minifilament
eruptions.  \citet{panesar.et16a} found that magnetic cancelation occurred at the coronal-jet location before and during the coronal jet in 10
quiet Sun jets, and in a similar study \citet{panesar.et18a} found magnetic cancelation occurring near the start of 13 coronal 
hole jets.  Additional multiple-coronal-jet studies support these findings, including \citet{mcglasson.et19} and 
\citet{muglach21}.  A study of 
\citep{kumar.et19} argues that ``shearing and/or rotational photospheric motions" are more important than cancelation
in producing coronal-hole jets that they studied; they do find, however, evidence that minifilament eruptions produce 
about two-thirds of the coronal jets, and that small-scale eruptions without cool-minfilament material cause the remaining ones.  
Magnetic cancelation has also been found to accompany many active region jets (\citeauthor{sterling.et16b}~\citeyear{sterling.et16b}, \citeauthor{sterling.et17}~\citeyear{sterling.et17}; \citeauthor{mulay.et16}~\citeyear{mulay.et16} state that cancelation, flux emergence, or 
cancelation-plus-emergence produce coronal jets they observe).  Single-event studies have also found cancelation to 
accompany coronal jets in many cases.  There are several other observational examples of magnetic flux cancelation leading 
to minifilament eruptions that produce coronal jets \citep[e.g.,][]{solanki.et19,yang.et19,mazumder19}.  See the 
above-mentioned reviews for additional citations.

Observations support that flux cancelation at the neutral line labeled ``B" in Figure~\ref{fig:s15_zu}(a) often triggers the minifilament
to erupt.  Similarly, magnetic flux cancelation also likely often builds the minifilament prior
to eruption, in some cases hours to $\sim$two days prior to its eruption \citep{panesar.et17}.  If the minifilament has
twist on it, perhaps supplied when canceling magnetic elements themselves contain shear derived from photospheric 
motions, that twist can be supplied to the coronal-jet's spire upon eruption of that minifilament and its reconnection with the 
coronal field, explaining why many coronal-jet spires display a spinning motion  during their evolution \citep[e.g.][]{moore.et15}.

\section{Some Coronal-jet-observing Instruments}
\label{sec-instruments}

We briefly introduce instruments often used for coronal-jet studies over recent decades.  Here we describe those instruments that are 
referred to most in this paper.  See the reviews listed in \S\ref{sec-introduction} for discussions of other instruments used in coronal jet 
observations.

In SXRs, coronal jets were first extensively observed with the \yohkoh/Soft X-ray Telescope \citep[SXT;][]{tsuneta.et91}, which operated from 1991---2001.  It had a detector with square pixels of width $2''\kern-0.5em.\,455$, and variable 
cadence with the fastest being about 2\,seconds, although it often ran with much coarser cadence.  Its followup was the X-ray Telescope \citep[XRT;][]{golub.et07} on the \hinode\ satellite, launched in 2006, and still operating as of this writing.  It has pixels of width 
$1''\kern-0.5em.\,02$, and --- for observations most appropriate for coronal jets --- operated with a cadence of $\sim$30\,seconds.  
Both SXT and XRT imaged with variable field of view (FOV), although for cadences sufficient to observe coronal jets of 
$\sim$10-minute lifetime both instruments used a FOV smaller than that of the full solar disk.

With  \yohkoh/SXT,  a large percentage of observed coronal jets occurred in active regions, with very few seen in polar regions 
\citep{shimojo.et96}.   \citet{koutchmy.et97} did see some polar coronal hole jets with SXT, but only with relatively long 
exposures of 15 and 30\,seconds.  In contrast, coronal jets are very prominent and common in polar regions in \hinode/XRT 
observations \citep{cirtain.et07}.  As discussed in \citet{hinode.et19} (in the subsection on coronal jets), this difference in 
visibility between the two SXR-imaging instruments can be understood because the filters that 
see the coolest SXR plasma with \yohkoh/SXT had sensitivity that dropped off sharply below about 2\,MK, while 
the coolest filters of \hinode/XRT have good sensitivity to plasmas of down to just under 1\,MK\@.  From filter-ratio 
temperature studies, \citet{nistico.et11}, \citet{pucci.et13}, and \citet{paraschiv.et15} determined that polar-coronal-jet 
spires have temperatures of $\sim$1---2\,MK\@.  Therefore this could explain why they are easily visible in 
images from the cooler-temperature-detecting \hinode/XRT, but much-less visible or invisible in images from the hotter-temperature-detecting 
\yohkoh/SXT\@.

For EUV observations, although there were some earlier useful observations with the EUV Imager (EUVI) telescope on the
\stereo\ spacecraft \citep{nistico.et09,nistico.et10}, the results discussed in the present paper largely derive from the 
Atmospheric Imaging Assembly \citep[AIA;][]{lemen.et12} on the Solar Dynamics Observatory (\sdo) satellite,
which has been operational from 2010, and is still operational as of this writing.  It has a detector of $0''\kern-0.5em.\,6$ 
pixels, and regularly observes the entire solar disk with 12\,second cadence in seven EUV channels centered at 
304, 171, 193, 211, 131, 335, 94\,\AA, roughly in order from detectability of the coolest to the hottest plasmas for non-flaring
situations (the details of the ordering depend on the distribution of temperatures in the emitting plasmas, and also
some channels have good response in multiple temperature ranges; \citeauthor{lemen.et12}~\citeyear{lemen.et12}
gives the AIA response curves and principle ions contributing to each wavelength band). \citet{sterling.et15} found
that polar coronal hole jets are best visible in the first four of these channels, and that the hotter channels of 131, 335, 
and 94\,\AA\ added little new information.  Active region jets, which tend to be hotter than the polar coronal hole
jets, generally are well seen over a broader range of AIA channels. \citet{shimojo.et00} found active region jets to 
have temperatures 3---8\,MK, based on SXT filter-ratio methods, 
and later studies have also found active region jet temperatures in this range \citep{paraschiv.et22}.  

Several papers by
\citet{mulay.et16,mulay.et17a,mulay.et17b} use 
the EUV Imaging Spectrometer (EIS) on \hinode\ to undertake spectroscopic studies of coronal jets, and survey a broader 
temperature range than that of the SXR 
filter-ratio methods of the just-mentioned studies.  They report the bulk of the emission of active-region-jet spires to 
be of temperatures $\ltsim$1---2\,MK\@.  This is substantially lower than the active-region-jet temperatures from 
the filter ratios mentioned above, e.g., the 3---8\,MK of \citet{shimojo.et00}.  But this difference is likely due
to the nature of the plasmas that the respective instruments can detect.  EIS, being an EUV spectrometer, 
has spectral coverage of substantially cooler spectral lines than those contributing to the SXR emission.
Thus it is likely that there is a wide distribution of plasma temperatures in coronal jets, and -- not surprisingly 
-- the SXR telescopes preferentially detect the hotter plasmas in those coronal jet spires, and therefore yield higher 
temperatures for coronal jets than the bulk of the coronal-jet plasmas detected by EIS\@. 

Spectroscopy in the UV from \iris\ has provided valuable insight into coronal jets.  This includes studies finding rotational
(spinning) motion in coronal jets \citep[e.g.,][]{cheung.et15,liu.et18,schmieder.et22,ruan.et19}, confirming indications of such rotation 
from earlier observations in EUV \citep{pike.et98}.  These spectra also provide information on densities in coronal 
jets \citep[e.g.,][]{cheung.et15,mulay.et17a,panesar.et22}.  Moreover, the high resolution of the \iris\ slitjaw images can 
complement the EUV and SXR coronal-jet observations in, for example, zeroing in on the fine-scale structure and 
dynamics at the coronal-jet magnetic-source location in the photosphere \citep{sterling.et17}.

Magnetograms from the \sdo\ Helioseismic and  Magnetic Imager \citep[HMI;][]{scherrer.et12} is frequently used to study
the photospheric magnetic flux values and changes around the base of coronal jets.  HMI has pixels of $0''\kern-0.5em.\,5$,
and takes a line-of-sight magnetogram of the full solar disk once every 45\,seconds.

\section{Coronal Jets and Jet-like Activity on Different Size Scales}
\label{sec-scales}

\subsection{Coronal-jet Physics on Large Scales}
\label{subsec-large_scales}

Coronal jets appear to be small-scale versions of larger eruptions, with the eruptive process that results in a minifilament eruption 
that produces a coronal-jet spire and a JBP corresponding to large-scale eruptions that result in filament eruptions and typical 
solar flares \citep{sterling.et15}.  That is, just as ``typical" solar filaments erupt (in what we are here calling ``large-scale
eruptions") to make long-observed ``typical" solar flares and that sometimes expel coronal mass ejections into the heliosphere, 
coronal-jets appear to be made by a minifilament eruption (a scaled-down version of a large-scale filament eruption) that leaves 
in its wake a JBP (a scaled-down version of a typical solar flare), and to result in material and magnetic disturbances that flow 
out along a spire and that sometimes flow into the heliosphere.

If coronal jets are indeed a scaled-down version of larger ``standard flare model" solar eruptions,
then we would expect other aspects of the smaller-scale eruptions that cause coronal jets and JBPs to have counterparts in the 
larger-scale eruptions  
that cause CMEs.  Here we discuss examining large-scale eruptions based on what has been 
found in coronal jets.

A characteristic of coronal jets is the anemone magnetic setup, similar to that shown in Figure~\ref{fig:s15_zu}(a).  There are many examples
of flares occurring from anemone active regions \citep{asai.et09,lugaz.et11,kumar.et13,devi.et20}.   \citet{joshi.et17} showed
that the setup for a large-scale eruption matched that of the coronal-jet minifilament-eruption picture, and that the dynamic motions
of the eruption matched closely that of an erupting minifilament producing a coronal jet.   A similar schematic was in fact 
drawn to explain a series of recurring solar eruptions much earlier (\citeauthor{sterling.et01}~\citeyear{sterling.et01},
\citeyear{sterling.et01a}, \citeyear{sterling.et01b}; these schematics in fact helped inspire the Fig.~\ref{fig:s15_zu} schematic
of \citeauthor{sterling.et15}~\citeyear{sterling.et15}).  These setups show that the same type of magnetic setup 
appears to be capable of generating similar solar expulsions both on the coronal-jet size scale, and on the size scale of typical 
solar eruptions.  Whether a coronal jet results or a CME results depends on 
how much of erupting minifilament/flux-rope lobe remains after the external reconnection in Figure~\ref{fig:s15_zu}(b) 
and~\ref{fig:s15_zu}(c). If the flux rope is robust enough to survive that reconnection (that is, if only the outer portion
of the flux-rope lobe is eroded away by external reconnection), then the remaining lobe and flux rope can escape to form a 
CME that carries a magnetic flux rope in its core region.  If on the other hand the external reconnection totally reconnects the flux rope, so that the
field lines that were previously closed in a flux rope all become open, then the feature becomes a coronal jet instead of 
a CME\@.  


An anemone setup appears to be necessary for coronal-jet formation, and formation of coronal jets in such a setup is supported
by numerical simulations \citep{wyper.et17,wyper.et18a,wyper.et18b,wyper.et19,doyle.et19}.  But large-scale eruptions
also occur outside of an anemone setup, and so we might ask whether the eruptions of minifilaments that cause coronal jets might 
also have similarities to larger-scale eruptions, independent of whether those larger-scale eruptions occur in an 
anemone configuration.  One possible such similarity is in the manner in which the minifilament eruptions and the large-scale
eruptions are triggered to erupt.  We have seen above that coronal-jet-producing erupting minifilaments
are apparently often built-up and triggered to erupt by magnetic-flux cancelation.  What about larger-scale eruptions?

\begin{figure}[h!]
\begin{center}
\includegraphics[angle=270,width=18cm]{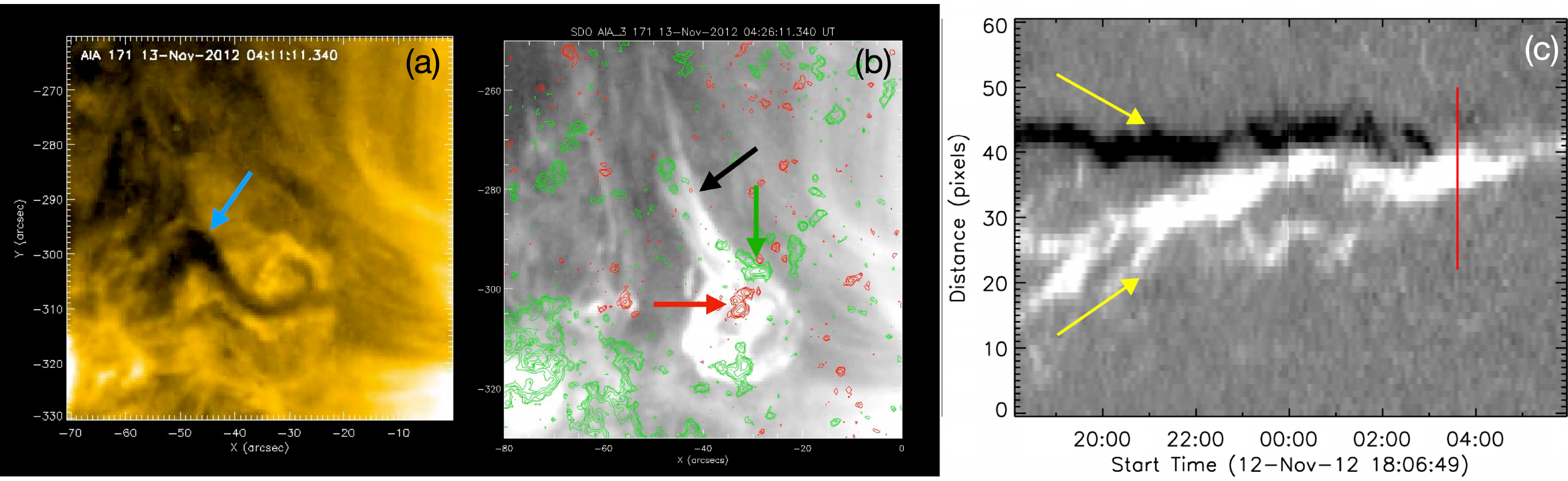}
\end{center}
\caption{Observations of a coronal jet originating from an on-disk location, on 2012 November~13, from \citet{panesar.et16a}.
Panel~(a) is an \sdo/AIA~193\,\AA\ image, showing a minifilament (blue arrow) starting to erupt to produce the coronal jet.  
Panel~(b) shows an AIA~193\,\AA\ image (grey scale) 15~minutes later, where the minifilament
has erupted and generated the coronal jet, with the black arrow pointing to the spire.  Overlaid is magnetic flux from an
\sdo/HMI magnetogram, with red/green contours indicating positive/negative polarities.  Red and green arrows point
to two polarity elements highlighted in the next panel.  Panel~(c) shows the time evolution of those positive-polarity (white) 
and negative-polarity (black) fluxes in a time-distance map, with the vertical red line indicating the time of coronal-jet occurrence.
This shows the polarities converging and undergoing cancelation near the time of the coronal-jet's onset.}
\label{fig:pan16_zu}
\end{figure}

\begin{figure}[h!]
\begin{center}
\includegraphics[angle=270,width=16cm]{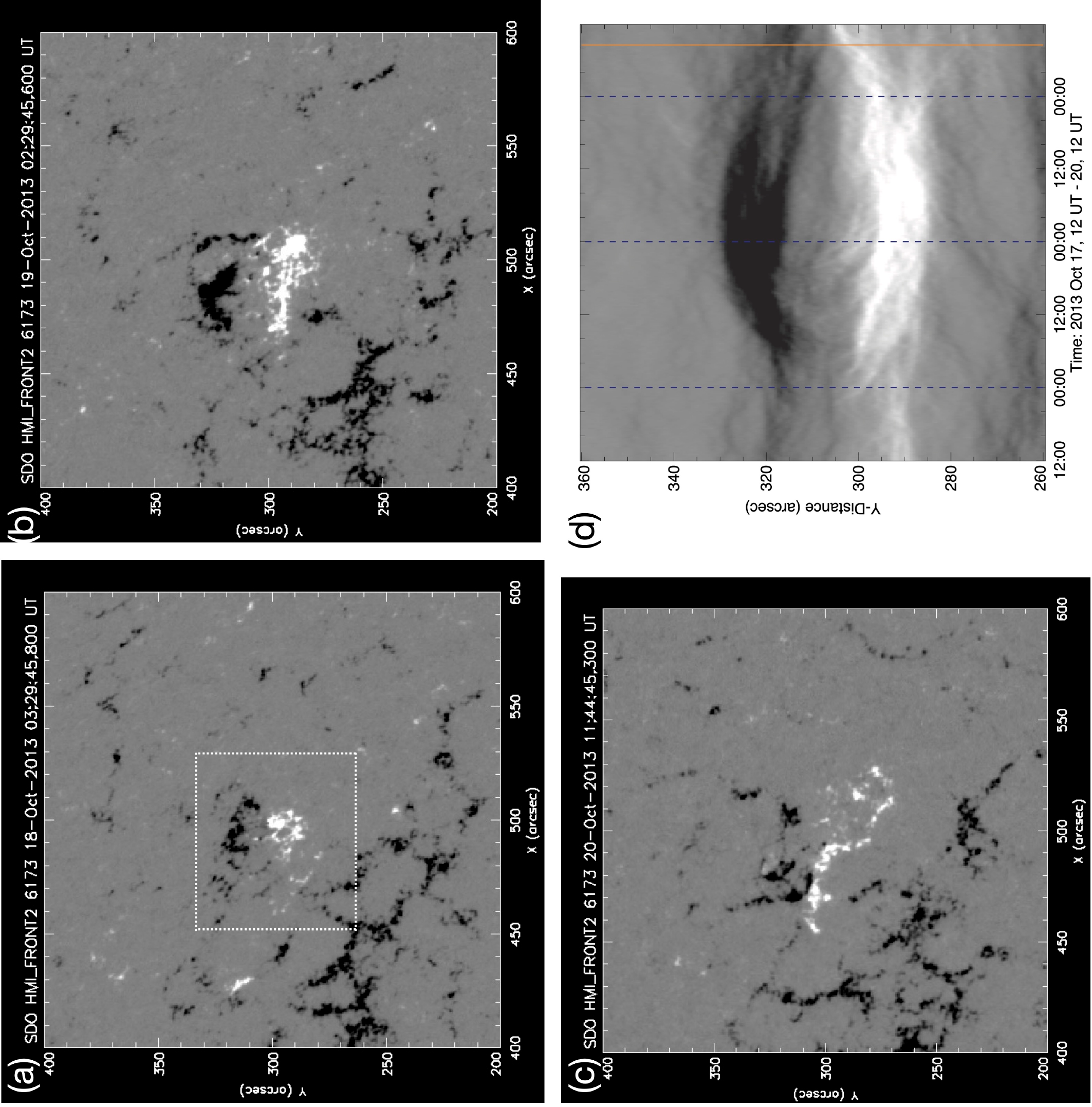}
\end{center}
\caption{\sdo/HMI magnetograms showing the evolution of a small-sized active region, NOAA AR 11868, which 
started emerging on 2013 October~16, and which produced a C-class flare and expelled a CME four days later 
on October~20. (a) The boxed region shows the emerging bipolar region about two days after the start of emergence.
Black/white indicate negative/positive magnetic polarities.  (b) The main negative and positive patches of the emerging
region continue to grow and spread apart.  (c) Portions of the opposite-polarity patches have converged on each other, and 
mangnetic cancelation is occurring between them.  (d) A time-distance plot 
showing evolution of the two polarity patches of the region.  This plot is formed by summing fluxes in the 
horizontal direction across the box in (a), and the y-axis of this panel is a sub-portion of the y-axis of
panels (a---c).   (From \citeauthor{sterling.et18}~\citeyear{sterling.et18}.)}
\label{fig:s_et18_event1_b_zu}
\end{figure}

To investigate whether large-scale eruptions are built up and triggered to erupt in a manner similar to coronal-jet-producing 
minifilament eruptions, \citet{sterling.et18}
studied how large-scale eruptions evolve toward eruption.    
In the case of coronal jets, the magnetic
elements taking part in the cancelation typically converge toward each other over the hours prior to the
eruption onset, as discussed in several papers \citep{panesar.et16a,panesar.et17,panesar.et18a,sterling.et17}, and 
as exemplified by Figure~\ref{fig:pan16_zu}(c).  This time period is short enough for those elements
to have relatively little interaction with surrounding flux elements.  In contrast, large-scale eruptions often occur in active regions that develop
for many days, or even weeks, prior to expelling an eruption (complex regions, such as delta regions, can evolve faster than
this, but the objective of \citeauthor{sterling.et18}~\citeyear{sterling.et18} was to compare with more standard 
eruptions).  Thus, in order to follow the region from the time of
flux emergence through to the time of the eruption, it was necessary to look at regions that were small enough for this
evolution to occur during a single disk passage of the region.  \citet{sterling.et18} presented two examples of this class.
In both cases the active regions were comparatively small bipolar active regions (total flux in each $\ltsim$10$^{21}$\,Mx).  Also in 
both cases the eruptions occurred about five days after emergence, and those eruptions produced CMEs observed in 
coronagraphs.  One of the regions remained almost completely isolated from any surrounding substantial flux over this period.
And the second (shown in Fig.~\ref{fig:s_et18_event1_b_zu}) was largely isolated from surrounding flux, although one of its 
polarities did have some interaction with nearby pre-exisiting opposite-polarity flux.

Both regions displayed similar evolution.  Figure~\ref{fig:s_et18_event1_b_zu} shows the evolution of one of these regions.
The boxed region in (a) shows a bipolar active region that is still emerging in this frame.  In panel~(b) the emergence 
is continuing, with centroids of the the main positive-polarity (white) and negative-polarity (black) patches further separated from
their central neutral line than they were in~(a).   By the time of~(c) however, they are no longer separating, and some of the 
opposite-polarity portions of the region have converged toward the central neutral line.  Panel~(d) shows a time-distance
map of this region, analogous to that in Figure~\ref{fig:pan16_zu}(c).  This shows that the polarities initially separate for
about one day following their initial emergence.  Their mutual directions then reverse, and the polarities start to converge.  
The orange line shows on this plot the time of the CME-producing eruption; this did not occur until after the polarities had 
converged on each other, and were undergoing flux cancelation along their central neutral line.  There were no CME-producing 
eruptions
from this region prior to this time.  \sdo/AIA observations show that the bright centroid of the resulting \goes\ C-class flare was on
that neutral line. Thus, similar to the situation with coronal jets, a flux rope eruption occurs along a cancelation neutral line.  In this
case, the region evolved for about four days with essentially no activity, and then had an eruption only after 
that cancelation started taking place. 

A second region examined in \citet{sterling.et18}, which also was substantially isolated magnetically from surrounding structures, 
began with flux emergence, underwent flux-polarity separation, and then had flux convergence and apparent cancelation along
its central neutral line among some portion of its two opposite-polarity patches.  This resulted in a eruption that produced a 
\goes\ B-class flare on the region's central neutral line (although in this case a second, weaker, eruption also occurred on a
neutral line formed from one of the emerging polarities and a pre-existing opposite polarity patch), and in the expulsion 
of a CME\@.  

\citet{chintzoglou.et19}, studying more complex magnetic situations involving multiple active regions, also found that eruptions 
occurred at flux-cancelation neutral lines.  

Returning to the discussion of the size scale of the erupting minifilaments that can cause coronal jets, \citet{moore.et22} examined
the evolution of 10 bipolar ephemeral active regions (BEARs) in a manner similar to \citet{sterling.et18}, but where they 
tracked their regions from emergence to disappearance.  These 10 regions produced 43 small-scale eruptions in total,
and all of these eruptions occurred at a neutral line in which apparent flux cancelation was taking place.  This again supports
that the the physics that causes eruptions on the coronal-jet size scale is essentially the same as that which causes 
large-scale eruptions that produce typical solar flares and CMEs.

These observations strongly support that flux cancelation is often essential in the
magnetic build-up and triggering of both smaller-scale eruptions that cause coronal jets, and larger-scale eruptions 
that cause flares and CMEs.  This is fully consistent with the 
mechanism for the build-up of the non-potential energy required for eruption via flux cancelation, as suggested by
\citet{vanball.et89}.  Observations of these processes occurring on faster time scales in the smaller-size-scale jets, however,
helps to elucidate strategies for investigating the processes in the more-slowly developing larger active regions.

\subsection{Possible Extensions to Smaller Scales}
\label{subsec-smaller_scales}

From the preceding discussions, we have presented evidence that the same basic process leading to eruptions  
occurs on two size scales -- that of large-scale eruptions and that of coronal jets.  \citet{sterling.et16a} considered a possible extended
relationship to smaller size scales, by plotting the the size of a typical filament or filament-like structure that erupts 
on one axis, and, on the other axis, a measure of the number of the respective eruptive events occurring at any given time 
on the Sun.   Their 
motivation was to see whether a substantial number of similar features might occur on a spicule 
size scale, assuming that the coronal-jet mechanism continues to scale downward to sufficiently small sizes.   If so, then it might 
be that at least some spicules (and perhaps many or most spicules) result from erupting filament-like features 
(erupting flux ropes) of that size scale.  (See \S\ref{subsec-spicules} for
a summary of spicule properties.)

The larger of the two size scales of eruptive events that have observed filament eruptions are the 
``typical" filament eruptions that have solar flares occurring beneath them, and -- in the case of ejective 
eruptions -- expulsion of a CME\@.  \citet{sterling.et16a} took a typical eruptive filament size to be 70,000\,km, 
with an appropriate scatter based on observed values
for a large number of filaments by \citet{bernasconi.et05}.  For the number of large-scale eruptions, they took 
observed CME rates  of from less than one to a few per day \citep{yashiro.et04,chen11}.  For coronal jets, they
estimated corresponding numbers for the size of erupting minifilaments based on measurements in \citet{sterling.et15},
which on average was just over 5000\,km for the erupting-minifilament lengths.  For the frequency of eruptions they
extrapolated rates given in \citet{savcheva.et07}, which yields about a few hundred per day over the entire Sun.

In order to compare with spicule occurrence rates, \citet{sterling.et16a} converted the occurrence rates for the large-scale 
eruptions and coronal-jet-size-scale
minifilament eruptions to the expected number of events occurring on the Sun at any random given time.  The motivation for
these units was to utilize the historical studies of spicule counts, which were sometimes expressed as the total number of spicules
on the Sun seen at a given time.  Those resulting values vary substantially \citep{athay59,lynch.et73}, but overall they are roughly in the 
neighborhood of $10^6$ spicules on the Sun at a given time. (\citeauthor{judge.et10}~\citeyear{judge.et10} estimate about
a factor of ten higher; see \citeauthor{sterling.et20c}~\citeyear{sterling.et20c}.)  

In order to complete the comparison with the larger erupting features, a value for the typical size of the erupting
filament-like flux rope that would produce a spicule is required.  No such cool-material eruptions have been convincingly 
observed to date, and therefore spicules being formed by the coronal-jet-producing minifilament-eruption-type 
mechanism is wholly speculative at this
point.  But because coronal-jet-spire widths are similar to the measured lengths of the erupting minifilaments that produce them,
by analogy, \citet{sterling.et16a} hypothesized that there might be erupting {\it micro}filaments (erupting micro flux ropes) 
of lengths comparable to the width 
of spicules (a few hundred km), that produces some spicules.

\begin{figure}[h!]
\begin{center}
\includegraphics[angle=0,width=11cm]{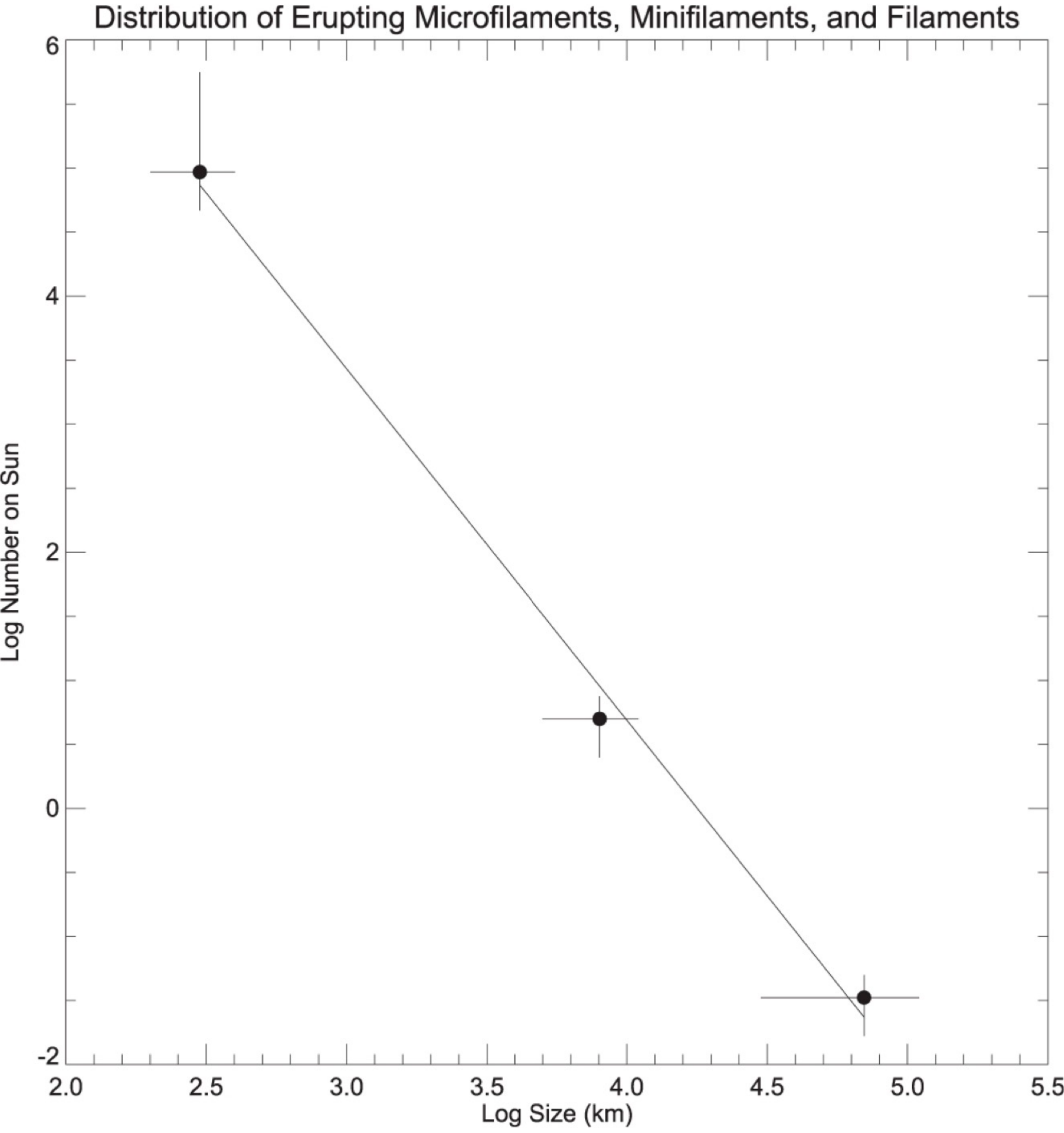}
\end{center}
\caption{Plot showing the instantaneous number distribution of filament-like eruptive events on the Sun, 
where those eruptions plausibly all operate by
the same general eruptive mechanism as that for coronal jets pictured in Fig.~\ref{fig:s15_zu}. The horizontal axis is the
size of the erupting cool-material feature, and the vertical axis shows an estimate of the number of eruptions 
of those features occurring on the Sun at any given time.  There are only three points plotted.  The 
right-most point represents large-scale eruptions of typical-sized filaments that produce solar
flares, and CMEs in the case of ejective eruptions.  The middle point represents coronal jets, such as
shown in Figs.~\ref{fig:s22_jet14} and~\ref{fig:pan16_zu}, and which are made by minifilament eruptions 
as depicted in Fig.~\ref{fig:s15_zu}.  
The leftmost point is for speculative ``erupting microfilaments" (erupting microfilament-carrying flux ropes) 
of the size scale of 
spicules, and which might produce spicules or spicule-like features in the manner of Fig.~\ref{fig:s15_zu}.  
``Error" bars represent uncertainties
in the observational parameters.  This plot demonstrates the plausibility that spicules might fit on the 
same power-law distribution as large-scale eruptions and the minifilament eruptions that make jets, but much
more evidence is required before a firm conclusion can be make regarding whether some or most spicules are
generated in a manner analogous to how coronal jets are made.
This plot is from \citet{sterling.et16a}; see that paper for further details.}
\label{fig:s_m16_zu2}
\end{figure}

Figure~\ref{fig:s_m16_zu2} shows the resulting plot.  When a linear extension is made from the large-scale eruptions through
the coronal-jet-sized eruptions, and then extended down to the size scale of the putative erupting microfilaments, the ordinate's value
for their occurrence rate falls near the lower end of the estimated spicule-number counts.  This implies that at least some
portion of spicules might be scaled-down versions of coronal jets, formed by eruptions of microfilaments, or it could be that multiple
spicules are produced by a single such eruption.  

We next consider from an observational standpoint the suggestion that the coronal-jet-producing mechanism operates on size scales smaller 
than that of coronal jets.

\subsection{Jetlets}
\label{subsec-jetlets}

\citet{raouafi.et14} studied long and narrow transient features of a scale smaller than coronal jets, using \sdo/AIA images and HMI magentograms.
They called these features {\it jetlets}, due to their similarity to coronal jets, except having a smaller size (both width and length).  Jetlets are
smaller than coronal jets, but larger than chromospheric spicules.  \citet{raouafi.et14} found
the jetlets to occur at the base of coronal plumes, and they suggest that they (along with transient base brightenings) are the result of 
``quasi-random cancellations of fragmented and diffuse minority magnetic polarity."  Therefore, these features seem to be smaller
versions of coronal jets.  (``Plumes" are long and narrow features – first noticed during total eclipses – extending out to several solar radii in polar regions.  Compared to jet-like features, plumes are long-lasting, persisting for $\sim$day.  See, e.g.,
\citeauthor{poletto15}~\citeyear{poletto15}.)

\citet{panesar.et18b} examined jetlets with \iris\ UV and AIA EUV images, and HMI magnetograms.  They studied ten jetlets, and
found them to have lengths $\sim$27,000\,km, spire widths $\sim$3000\,km, base size of $\sim$4000\,km and speed $\sim$70\,\kms.
They argued that jetlets are a more general solar feature than presented in \citet{raouafi.et14}, occurring at the edges of chromospheric 
network both inside and outside of 
plumes.   In agreement with \citet{raouafi.et14}, they found that magnetic flux cancelation was the likely cause of the jetlets, just as
it often leads to coronal jets.  Furthermore, their jetlets were accompanied by brightenings analogous to the JBP seen at the base 
of coronal jets.  Based on these and other characteristics, \citet{panesar.et18b} concluded that the jetlets are likely scaled-down 
version of coronal
jets, and that they are consistent with the erupting-minifilament scenario for their production. They did not, however, observe a 
clear indication of the existence of an actual cool-material erupting minifilament at the base of their jetlets.  

\citet{panesar.et19} extended these studies to even smaller-sized jetlets, using EUV 172\,\AA\ (\feix/\fex) images from the
Hi-C\,2.1 rocket flight.  This instrument had pixels of width $0''\kern-0.5em.\,129$, and
cadence of 4.4\,seconds; see \citet{rachmeler.et19} for details.  Six events were identified from the data from Hi-C\,2.1's five-minute flight.
As with the \citet{panesar.et18b} jetlet study, these events also occurred at the edges of network cells.  On average they 
had spire lengths of $\sim$9000\,km, widths of 600\,km, and speeds of $\sim$60\,\kms.  At least four of these events
seemed consistent with being small-scale coronal jets following the erupting-minifilament mechanism, although once again
there were no direct observations of erupting cool-material minifilaments.

\subsection{Spicule-sized Features}
\label{subsec-spicules}

Spicules are chromospheric features that are extremely common, have lengths $\sim$5000---10,000\,km, widths of a few hundred 
km, lifetimes of a few minutes, and cluster around the magnetic network \citep[e.g.][]{beckers68,beckers72}.  Although many 
ideas exist, their generation mechanism is still unknown; see discussions in various review works \citep[e.g.,][]{beckers68,beckers72,sterling00,tsiropoula.et12,hinode.et19,sterling21}.    

As mentioned in \S\ref{subsec-smaller_scales}, erupting microfilament flux ropes have not been observed in spicules.  They may, however, be
present, but hard to observe for a variety of reasons.  Similarly, a bright point that corresponds to a JBP has not been convincingly 
observed at the base of spicules.  These points are not consistent with an erupting-microfilament mechanism for spicules.  Nonetheless,
these absences are not definitive evidence that these features do not exist in spicules; they may exist but be hard to detect, as 
discussed in \citet{sterling.et20c}.  

Moreover, there are several observations that are consistent with a microfilament-eruption mechanism for spicules.  One of these 
is the observation of mixed polarity elements at the base of many spicules.  Fresh evidence for this is presented in 
\citet{samanta.et19}, obtained using state-of-the-art ground-based observations in the pre-DIKIST era.  This work presents 
evidence that spicules result from dynamic activity at the base of spicules, 
which could be due to flux cancelation and/or emergence.  As discussed above, there is extensive evidence that many coronal 
jets result from
flux-cancelation episodes.  Ideas for coronal-jet production from flux emergence has also been presented \citep{yokoyama.et95}.

Spicules also display characteristics of spinning motions \citep{pasachoff.et68,depontieu.et12,sterling.et20b}.   We have noted that 
the minifilament-eruption idea offers an explanation for the spinning of coronal jets, when an erupting twisted minifilament 
transfers its twist to the coronal-jet-spire's coronal field via external reconnection.  Thus the same mechanism acting on speculative erupting microfilaments might explain this spicule spinning as magnetic untwisting.

There remains, however, the possibility that spicules are driven by any of a number of other suggested mechanisms (see the above-cited
reviews), and many of these ideas cannot yet be ruled out.  Spicule-sized features that work via the coronal-jet-production 
mechanism may instead 
drive other features of that size, such as the UV network jets described by \citet{tian.et14} \citep[some of which are UV observable 
in EUV, at least in AIA 171\,\AA\ images;][]{tian.et14},
or the ``chromospheric anemone jets" observed in active regions, or similar jet-like features observed in plages \citep{depontieu.et04,sterling.et20c}.

\section{The Importance of Coronal Jets}
\label{sec-importance}

Coronal jets are important for solar physics in a number of ways.  One of these is to gain insight into the buildup and onset of 
large-scale eruptions.  Studies of these eruptions has importance in a variety of areas, ranging from the understanding 
of key inputs to Space Weather to gaining insight into fundamental astrophysical processes.

One of the key unknowns of solar physics is the details of the mechanism leading up to and causing large-scale 
solar eruptions that produce typical-sized-filament eruptions, solar flares, and CMEs.  The revelation that many coronal 
jets are scaled-down versions of large eruptions has important implications for resolving these questions.  Not only are 
coronal jets of a smaller size scale than those large eruptions, but also the pre-coronal-jet evolution time scale of coronal 
jets is substantially shorter than
that for large eruptions: Large eruptions from bipolar active regions usually require at least many days to build up the magnetic circumstances
that result in the eruption \citep[e.g.,][]{sterling.et18}, while in contrast, for a sample of quiet-region jets the corresponding time 
scale was found to be of the order of hours or a couple of days \citep{panesar.et17}.  Additionally, it is difficult to find examples 
of magnetically isolated large-scale regions that produce large eruptions for many days while they remain on the Earth-facing side of 
the Sun, and this exacerbates the difficulty in unraveling the fundamental processes that lead to eruptions.  On the other hand,
it is often easy to find and follow coronal-jet-producing regions on the Sun, by working backwards from the time of coronal-jet occurrence.
As discussed in \S\ref{subsec-large_scales}, \citet{sterling.et18} used these points to learn about regions leading to flares and CMEs in 
two small active regions, largely based on lessons learned from coronal-jet studies.  These studies and comparisons are,
however, still vastly incomplete.  Careful study of the details of eruptions happening on the coronal-jet-sized scale are necessary to
understand fully the coronal-jet-production process.  And then the findings can be used as starting input for improved studies of
large-scale events.

And as demonstrated in \S\ref{subsec-smaller_scales}, coronal jets almost certainly can provide insight into the operation of some 
smaller-scale events.  This is in particular true for those smaller
objects with the most apparent coronal-jet-like qualities; this includes jetlets, at least some of which must be scaled-down versions
of coronal jets.  It will be important to see how far the obvious similarities continue down in size scale by using long-term high-resolution
and high-cadence observations at appropriate wavelengths.  This will help determine whether the similarities continue down to the
size of the UV network jets, which are nearing spicule size.  Because the solar wind appears to originate from most if not all locations 
on the Sun (in particular, the fast solar wind originates from open-field areas, while the slow wind comes from 
closed-field regions), the solar-surface events that drive it are likely distributed 
almost uniformly over the surface.  An understanding of jetlets and UV network jets will provide insight into whether these events plausibly 
power the solar wind.   Careful additional coronal-jet studies are required to guide and assist observational investigations of these yet 
smaller-scale features.

Coronal-jet physics may have even broader implications beyond the most obvious looking coronal-jet-like or solar-eruption counterparts.  For example,
\citet{katsukawa.et07} have examined ``penumbral jets," rooted along the filament-like low-lying magnetic loops (the penumbral fibrils) 
that spread radially from sunspot
umbra and form the penumbra \citep{tiwari.et13}.   \citet{tiwari.et16,tiwari.et18} also studied penumbral jets, and found that large ones
originated from the end of the penumbral filament
rooted farthest away from the umbra, and that these regions had mixed magnetic polarities that underwent magnetic cancelation 
near the time of
penumbral-jet generation.  Moreover, using \iris\ spectra, they found Doppler evidence that the penumbral jets are undergoing spinning motion.
Although these penumbral jets are $\sim$100 times smaller than typical SXR coronal jets, their properties of magnetic cancelation 
at their bases, long (compared to their width) spires, and spin, are similar to what has been found in numerous coronal
jets.  This suggests that the basic physical mechanism creating these penumbral jets might be essentially the same as that which produces 
coronal jets.

``Campfires" are features recently discovered in high-resolution EUV images from the Extreme Ultraviolet Imager (EUI) 
on the \so\ spacecraft, appearing as small localized brighteings of size scales of a few 100---few 1000\,km and lasting 10---200\,seconds \citep{berghmans.et21}.   By comparing a selection of campfires with \sdo/HMI magentograms, \citet{panesar.et21} present evidence
that they occur on canceling magnetic neutral lines, similar to how coronal-jet-producing erupting minifilaments and accompanying base 
brightenings also frequently occur on canceling neutral lines.  This suggests that the same process that makes coronal jets (small-scale-filament/flux
rope eruptions) might also make campfires.    

If the same processes indeed make features as varied as large-scale solar eruptions, coronal jets, jetlets, some spicules, UV network jets,
campfires, penumbral jets, and perhaps other features, then it is important for heliophysics to clarify which magnetic and physical 
circumstances produce which feature in what situations.

Coronal jets also have influence far out into the heliosphere.  White-light coronagraph images show that some CMEs are relatively narrow,
with angular extents $\ltsim$5$^\circ$; these features have been called both ``narrow CMEs" and 
``white-light jets" \citep[See discussion and references in][]{sterling18}.    \citet{wang.et98}
have shown that these features often originate from jetting activity at the solar surface, and a mechanism for producing these
features has been presented, based on the minifilament-eruption picture, by \citet{panesar.et16b}.  \citet{sterling.et20a} present
observations of material expelled from coronal jets extending out to tens of solar radii in images from the  \stereo\ ``Hi1" Heliospheric Imager. 
There is also evidence of {\it in situ} detection of coronal-jet material in the solar wind \citep{yu.et14,yu.et16}.
Both \citep{sterling.et20a}, and a followup work \citep{neugebauer.et21}, suggest that coronal jets, and/or smaller jet-like 
features that work via the minifilament-eruption mechanism, might propagate out to the heliospheric locations of the  \psp\ satellite,
and be detected as magnetic \al ic kinks in the field that are known as ``switchbacks" \citep{bale.et19,schwadron.et21}.  The idea is that the erupting minifilaments might carry 
twist, and external reconnection with the ambient coronal field could transfer that twist to the field, as described by \citet{shibata.et86}.
That twist could convert to swaying of inner coronal field \citep{moore.et15}, and then steepen into an \al ic kink -- forming the switchback 
-- due to variations of the \al\ speed in the solar wind \citep{sterling.et20a}. Switchbacks are extremely common in the solar
wind at distances of a few tens of solar radii \citep{bale.et19}.  They also appear to carry the imprint of size scales at the Sun 
corresponding to supergranules \citep{bale.et21,fargette.et21}, and even of granules \citep{fargette.et21}.  The minifilaments that 
erupts to make coronal jets \citep[$\sim$10,000---few $\times$10,000\,km; e.g.,][]{sterling.et15,panesar.et16a} are not so 
different from supergranule scales ($\sim$40,000\,km), while the width of spicules and similar features (few 100\,km or so) is 
not too different from the size scale of granules ($\sim$1000\,km).  Better observations of coronal jets can help determine whether this
switchback-production idea matches detailed observations, or whether a different mechanism might be responsible for the
switchbacks, such as ``interchange reconnection" ideas \citep[e.g.,][]{zank.et20,owens.et20,drake.et21,schwadron.et21}, or any 
of several ideas for generating the switchbacks in the solar wind \citep[see citations in][]{fargette.et22}, while keeping in mind that
the mechanism should explain the observations of a supergranule- and/or granule-size-scale dependence of switchback size scales.

These points illustrate the value to solar physics -- and beyond -- of understanding the nature of the processes that lead up to
and produce coronal jets.  To this end, we suggest a new instrument focused on observing coronal-jet-sized features in the low
corona.  We first note, however, that another reason for further observations of coronal jets is to learn more about the nature of 
coronal jets themselves.   Even though there is strong evidence that the minifilament eruption model explains many coronal jets,
it is still to be confirmed that the model holds up to close scrutiny under improved observations.  An alternative mechanism,
which was originally proposed along with the earliest detailed coronal-jet observations \citep{shibata.et92}, is that they result
when emerging magnetic flux reconnects (external reconnection) with surrounding coronal field.
As mentioned in \S\ref{sec-introduction}, many observations fit the minifilament-eruption model and find that coronal jets frequently
occur as the result of minifilament (or flux rope) eruptions on canceling magnetic neutral lines.  Other observations
of coronal jets -- such as the direction of motion of coronal-jet spires horizontal to the solar surface -- are also consistent wtih the minifilament-eruption
mechanism and not with the emerging-flux mechanism \citep{baikie.et22}.  So questions are: do coronal jets ever form via
the emerging-flux mechanism, and if so, are there any special characteristics of those coronal jets compared to those that we
have described in \S\ref{sec-introduction}?  Another question is, if coronal jets do not occur frequently via the emerging-flux 
mechanism, then why not?  After all, numerical simulations indicate that flux emerging into open surrounding coronal
field should produce a coronal jet \citep[e.g.,][]{yokoyama.et95,nishizuka.et08}.  So if that process does not produce coronal jets on the
Sun in reality, then it is important to understand why the reconnection resulting from flux emergence (which inevitably 
must happen) does not produce actual coronal jets.  These questions emphasize the importance of understanding physics
on the size scale of coronal jets.

\section{The ``SEIM" Instrument to Observe Coronal Jets and Jet-like Structures}
\label{sec-seim}

We propose a new instrument for the next generation, under the provisional name of the ``Solar Explosions IMager" (SEIM).\@
This instrument would be tuned to observe features of size scales of coronal jets at EUV wavelengths, with
a spatial resolution and cadence similar to that of the Hi-C2.1 instrument. The Hi-C flights, however, were on sounding rockets, and
so of short duration ($\sim$5\,min).  The \so/EUI instrument can also achieve high resolution comparable to that of Hi-C, but such high
resolution is only available for a few days around \so's perihelion.   Our idea is for SEIM to be on a satellite, allowing for 
long-term high-resolution observations.   We now outline the instrument's desired characteristics.

\subsection{Wavelength Coverage}
\label{subsec-wavelength}

As pointed out in \S\ref{sec-instruments}, coronal jets in polar coronal holes such that the spires are generally seen in the \sdo/AIA 
channels of 304, 171, 193, and/or 211\,\AA.\@    Among these, 171, 193, and 211 are all coronal lines.  It would be best to
include all of these channels, as sometimes the spire tends to be better seen in one than the other.  Nonetheless, if it is
visible in one of these three channels, it is usually at least detectable in the other two, based on observations of 41 coronal jets in
\citet{sterling.et15} and \citet{sterling.et22}.  Therefore, as a minimum, one of these three channels should be included in a
minimal mission.   The 304\AA\ channel shows a mixture of what might be called ``upper"-chromospheric 
and transition region plasmas.. It sometimes shows features of coronal jets detected in SXRs that are not apparent in the other 
three cool-coronal (171, 193, 211\,\AA) channels \citep{sterling.et22}, and therefore that channel would be
essential to include in a SEIM mission.
 
Brighter coronal jets, such as those occurring at the periphery of active regions, are often visible in all AIA EUV channels.  Therefore,
an instrument designed to see coronal-jet-like features in coronal holes (and quiet Sun) well would also be able to see coronal jets and 
similar features in active regions.

A channel that shows photospheric emissions, such as AIA's 1600\,\AA\ channel, would be essential to facilitate comparisons with
other instruments such as DKIST\@.  This channel also shows ribbon-like flare emission at the base of some active-region jets
\citep{sterling.et16b}.

Given these considerations, a minimal wavelength-coverage package for a SEIM instrument could be 304, 171 or 193, 94, and 1600\,\AA\@.

\subsection{Resolution, Cadence, and Field of View}
\label{subsec-other_parameters}

In active regions, some strands of erupting minifilaments are substantially thinner than the width of erupting minifilaments in 
coronal holes or quiet regions, having widths of $\ltsim$2$''$ \citep{sterling.et16b}.  Hi-C, either in its original incarnation \citep{kobayashi.et14}
or the Hi-C2.1 version discussed above (\S\ref{subsec-jetlets}), would be able to resolve many of these, and in general sees 
features at the limit of or beyond what is readily detectable in AIA \citep[e.g.,][]{brooks.et13,tiwari.et16,panesar.et19,tiwari.et19,sterling.et20c}.
Based on this, we are confident that a resolution comparable to that of Hi-C ($0''\kern-0.5em.\,1$ pixels) will be adequate for revolutionary 
breakthroughs in the study of coronal jets and similar-sized phenomena.

For coronal-jet studies, the 12-second cadence of AIA has been adequate.  Smaller-scale jet-like features can have lifetimes shorter than the
$\sim$tens of minutes of coronal jets, including $\sim$one minute for small jetlets \citep{panesar.et18b} and UV network jets.
Therefore a faster-than-AIA time cadence comparable to that of Hi-C, about 5\,seconds, would be preferred in order to sample 
these objects well.

\begin{figure}[h!]
\begin{center}
\includegraphics[angle=0,width=12cm]{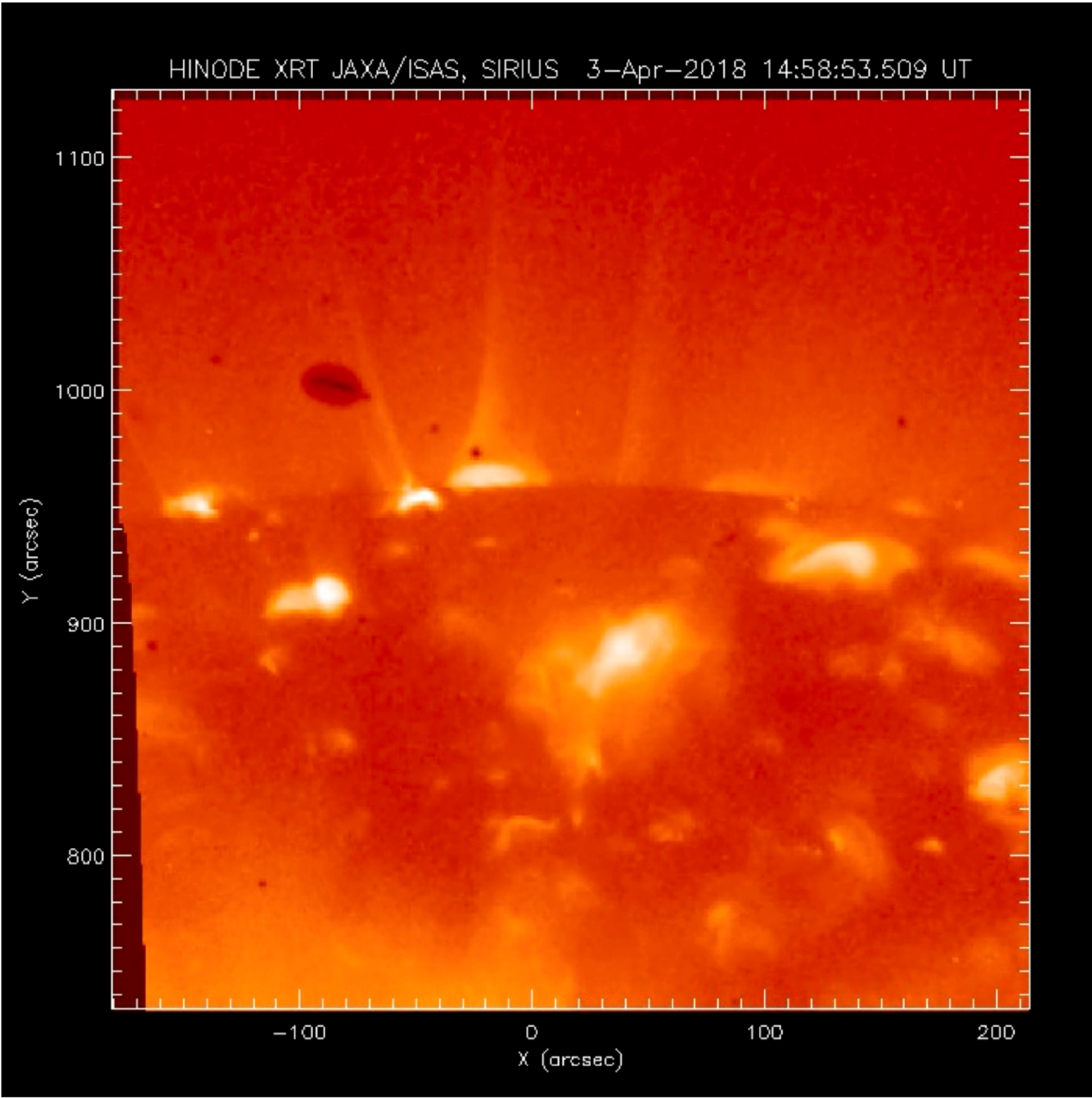}
\end{center}
\caption{\hinode/XRT image in SXRs  \citep[from][]{sterling.et22} of the northern polar coronal hole region, showing 
several coronal jets.  The field of view (FOV) here is $6'\kern-0.5em.\,.67 \times 6'\kern-0.5em.\,.67$.  This is a little larger than 
the minimum FOV ($6'\kern-0.5em.\,.0 \times 6'\kern-0.5em.\,.0$) that would be desired for
the suggested SEIM mission. (The dark oval near (-100,1000) and similar smaller dark spots are imaging artifacts.)}
\label{fig:s22_xrt4}
\end{figure}

A FOV of about $6' \times 6'$, would be acceptable for an initial mission.  This would be slightly larger than the 
$4'\kern-0.5em.\,.4 \times 4'\kern-0.5em.\,.4$ Hi-C2.1 FOV \citep{rachmeler.et19}.  \sdo/AIA's detector is a circle of 
diameter $41'$ \citep{lemen.et12}, and so our proposed detector would have a FOV of about one-sixth that of 
AIA's.  Thus we could obtain our goal of  $0''\kern-0.5em.\,1$ with a FOV one-sixth that of AIA's, by using an AIA-sized
detector ($4096 \times 4096$ pixels$^2$).  Advances in technology might make it feasible to improve upon this, allowing for 
increased FOV and/or higher
resolution, but these minimal criteria would allow for substantial advancements in our understanding of coronal jets and jet-like features.
Figure~\ref{fig:s22_xrt4} shows a sample image from XRT with a FOV similar to that being discussed (that image's FOV is only 
slightly larger, at $6'\kern-0.5em.\,.67 \times 6'\kern-0.5em.\,.67$), with much of the northern polar coronal hole and several 
X-ray coronal jets visible.

\subsection{Orbit, Accompanying Instrumentation, and Operations Planning}
\label{subsec-orbit}

Ideally, a mission carrying a SEIM instrument would have extended, uninterrupted views of the Sun.  Accordingly, a Sun-synchronous 
orbit, such as that of \hinode\ \citep{kosugi.et07} or \iris\ \citep{depontieu.et14b} would be appropriate, allowing for $\sim$nine mouths
of uninterrupted viewing, and $\sim$three months with orbits that include spacecraft nights while still allowing $\sim$one hour of solar
observing per orbit.  Longer periods of uninterrupted viewing 
would be possible from L1 or a similar location, but that might be more appropriate for a more extensive followup mission.

An imaging-only mission plan would be of limited value for advancing the science of jet-like features.  As a minimum, systematic
corresponding line-of-sight magnetograms would be essential to complement these observations.  This could be included on 
the same spacecraft, in which case a magnetogram FOV comparable to that of the EUV instrument would be acceptable.  Alternatively, 
it would be possible to use synoptic full-disk magnetograms from elsewhere if appropriate ones are available; for example, it would
be fully acceptable to rely on magnetograms from \sdo/HMI or a similar instrument on a different satellite that is operational at the 
time of a SEIM mission.    In either case, the time cadence should be 
about $\sim$1\,minute, comparable to that of \sdo/HMI (45\,seconds).  Spatial resolution of HMI's level would be the minimum desired, 
but would be adequate for an initial mission.  For special programs, coordinated observations with DKIST or other ground-based
instruments \citep[such as BBSO; e.g.,][]{samanta.et19} would allow for much higher-resolution magnetograms.  It will also be extremely
valuable to have spectroscopic observations at UV and/or EUV, or even SXR, wavelengths, to obtain diagnostic information on 
the observed objects.  These spectra should have sufficient spatial and spectral resolution and high-enough cadence to address questions
such as whether jetlet, UV network jet, and even spicule-sized objects routinely display characteristics of jets, such as spinning 
motion of their spires.

We know that coronal jets are common in polar coronal holes, and they are observed in on-disk coronal holes also.  Therefore,
a basic minimal-maintenance plan would be to observe (with tracking) an on-disk coronal hole when one is available.  This would
allow for coordination with line-of-sight magentograms.  A second low-latitude target would be active regions.  The frequency of 
typical coronal jets from active regions is not yet known, but they are not uncommon.  Even in the absence of such coronal jets, there
are smaller-scale penumbral jets that that are ubiquitous in active regions, and it would be desirable to have high-resolution, 
high-cadence observations of other active-region activity, and of course it would be highly desirable for the instrument to observe 
large-scale eruptions from active regions.  In the absence of on-disk coronal holes and active regions (or if there are only 
active regions showing essentially no substantial activity), one of the two polar regions (preferably one with a prominent coronal hole) 
would be the standard default target.

\subsection{Extensions}
\label{subsec-extensions}

The instrument proposed here could act as a proving ground for a more elaborate mission that features a full-disk FOV and
wider wavelength coverage.  This would be analogous to how the \trace\ mission \citep{handy.et99} preceded AIA on \sdo\@. 
It would be fully appropriate for such a more-extensive mission to operate from L1 or similar location, with uninterrupted solar
viewing.  Such an instrument would ideally be accompanied by a complementary magnetograph, and perhaps other instruments,
on the same spacecraft.

The Appendix provides a summary of properties of coronal jets and jet-like features that will be either observed directly with SEIM, or
to which SEIM will provide valuable supplementary observations, for refining our understanding of all of these features.

Even in the simplest form however, the SEIM instrument suggested above would be far more than just a “solar-jets telescope.”  
As we have argued above, solar jets can be viewed as a proxy for one of the many types of possibly similar solar features, on 
both larger and smaller size scales, that could be observed and studied in detail with such as instrument.  Therefore, SEIM would 
provide a new, high-resolution, high-time-cadence window into an understanding of fundamental explosive phenomena that occur 
on multiple size scales in the lower solar atmosphere, and that possibly power the heliosphere as well.

\section*{Conflict of Interest Statement}

The authors declare that the research was conducted in the absence of any commercial or financial relationships that could be construed as a potential conflict of interest.

\section*{Author Contributions}

\noindent
ACS collected all materials and authored the bulk of the manuscript.  RLM consulted on all aspects of the content.  NKP 
contributed consultations and input into coronal jet and jetlet details.  TS contributed consultations and input into spicule observation and analysis details.  SKT contributed consultations and input into sunspot and penumbral jet details.  SS critiqued the manuscript and
provided valuable feedback.

\comment{
The Author Contributions section is mandatory for all articles, including articles by sole authors. If an appropriate statement 
is not provided on submission, a standard one will be inserted during the production process. The Author Contributions statement 
must describe the contributions of individual authors referred to by their initials and, in doing so, all authors agree to be accountable 
for the content of the work. Please see  
\href{https://www.frontiersin.org/about/policies-and-publication-ethics#AuthorshipAuthorResponsibilities}{here} for full authorship criteria.
} 

\comment{
\section*{Funding}
Details of all funding sources should be provided, including grant numbers if applicable. Please ensure to add all necessary funding information, as after publication this is no longer possible.

THIS WILL BE HANDLED IN ACKNOWLEDGEMENTS
} 

\section*{Acknowledgments and Funding}

This paper expands upon material presented previously in a white paper submitted to the National Academies of 
Sciences, Engineering, and Medicine for the U.S. Solar and Space Physics (Heliophysics) Decadal Survey, for 
Heliophysics 2050 \citep{sterling.et22a}.

ACS, RLM, and NKP were supported with funding from the Heliophysics Division of NASA's Science 
Mission Directorate through the Heliophysics Supporting Research (HSR, grant No.~20-HSR20\_2-0124) Program,
and the Heliophysics Guest Investigators program.
ACS and RLM were supported through the Heliophysics System Observatory Connect (HSOC, grant No.~80NSSC20K1285) 
Program.  ACS received additional support through the MSFC \hinode\ Project, and NKP received additional 
support through a NASA SDO/AIA grant.  SKT gratefully acknowledges support by NASA HGI  
(80NSSC21K0520) and HSR (80NSSC23K0093) grants, and NASA contract NNM07AA01C (\hinode).  \hinode\ is a Japanese 
mission developed and launched by ISAS/JAXA, with NAOJ as domestic partner and NASA and UKSA as international 
partners. It is operated by these agencies in co-operation with ESA and NSC (Norway). 
We acknowledge the use of data from AIA and HMI data, both of which are instruments onboard \sdo, a mission of
NASA's Living With a Star program.

\section*{Data Availability Statement}
No new data were analyzed for this summary-type presentation.  All background material presented in the figures is available in
the references given in the corresponding figure captions.

\bibliographystyle{Frontiers-Harvard} 
\bibliography{ms}

\clearpage

\section{Appendix}

Here we summarize information about selected jet-like features in the solar atmosphere, and state how SEIM will
enhance our understanding of the mechanism(s) driving these features.  This listing is not intended to be exhaustive.

For each entry we provide the following. (1) The chief references for the numbers listed in the given entry.  Because many authors
and observers measure parameters in different ways, it is difficult to summarize accurately all of the values outside of an
extensive review article.  Therefore we list the sources used for presenting the specific values listed in this Appendix.  (2) Wavelength:
the principle wavelength regime where the features are observed.  (3) Appearance:  The general defining characteristic
appearance of  the features.  (4) Size:  Approximate characteristic size for the indicated dimension(s) and/or portion(s) of
the features. (5) Lifetimes: typical durations of the events.  (6) Energies.  (7) Locations:  Additional information of where on the Sun the
features are typically observed.  (8) SEIM Objectives:  Some of the unresolved aspects of the features and/or their production
mechanism to which SEIM can contribute improved understanding.

\noindent
{\bf Coronal jets:}  {\it Chief references for following parameters}: \citet{savcheva.et07,panesar.et16a,sterling.et18}.  
{\it Wavelength:}  mainly EUV and SXRs, but can have manifestations 
at other wavelengths too.  {\it Appearance:}  Has a spire that grows to be long and
narrow, emanating from a base region that is typically bright in SXRs.  {\it Size}:  spire; typically 
$\sim$10,000 $\times$ 50,000\,km in SXRs; base: up to a few $\times10^4$\,km in 
width in EUV\@.  {\it Lifetimes}: tens of minutes.  {\it Energies}: $\sim10^{26}$---$10^{29}$\,erg. {\it Locations}:  Exist in various solar 
regions, including coronal holes (sometimes called ``coronal hole jets"), quiet Sun (``quiet Sun jets"), 
and the edges of active regions (``active region jets").  {\it SEIM Objectives}: Many coronal jets result from eruptions of 
minifilaments, where the eruptions can be either confined or ejective, and where
the eruptions are often observed to result from magnetic flux cancelation.  A localized brightening (called the jet bright point or 
the ``JBP") usually forms on one edge of the base of the coronal jet, below the erupting minifilament.  The JBP is argued to be 
a small-scale version of a typical solar flare arcade that forms below a typical erupting filament.
Many coronal jets show spire rotations, consistent with magnetic untwisting.  See main 
text and Figure~\ref{fig:s15_zu} for more details on the proposed mechanism.  SEIM, in conjunction with other 
instruments, will test further this idea for the production of coronal jets, and investigate what 
percentage of coronal jets are consistent with this production mechanism, and how many might be due to a different
mechanism.

\noindent
{\bf Surges}:  {\it Chief references for following parameters}: \citet{foukal13,moore.et10,moore.et13,sterling.et16b,yokoyama.et95,th95}.  
{\it Wavelength}: primarily chromospheric lines, UV, and EUV~(304\,\AA)\@.    {\it Appearance}: 
traditionally seen as violent expulsions of chromospheric material from bipolar footpoints 
in active regions, with bright flare-like base brightenings.  {\it  Size}: maximum length: up to, but generally 
much less than, (1--2)\,$\times 10^5$\,km.   {\it Lifetimes}: few tens of minutes.  {\it Energies}: $\ltsim 10^{30}$\,erg. 
{\it Locations}: Traditionally in 
active regions.    {\it SEIM Objectives}: Although traditionally defined as an active region phenomenon, surge-like
events are seen in quieter regions also.  EUV~304\,\AA\ surge-like ejections are often seen coincident
with coronal jets, even in coronal holes.  SEIM will observe coronal-jet locations both in and outside of 
active regions, and determine whether they are the same basic physical phenomenon.  The connection 
between the surge-like cool-plasma ejecta and the hotter coronal-jet spires that they accompany can be 
explained in different ways, depending upon the mechanism assumed to be creating the coronal jets.  We expect
SEIM to see the cool surge material in absorption, and also the hotter bright 
base of many surges; these observations, in conjunction with other instruments, will help determine which idea for 
coronal-jet production most closely aligns with observations of surges and surge-like features.

\noindent
{\bf Jetlets}:  {\it Chief references for following parameters}: \citet{raouafi.et14,panesar.et18b,panesar.et19}.  
{\it Wavelength}: EUV, mainly 171, 172, and 193\,\AA; UV\@.   {\it Appearance}: spire that grows to be long and
narrow, similar to coronal jets but smaller in size. First identified at the base of plumes, but later found to be common more 
generally at the edges of the magnetic network. {\it Size}: maximum length: $\sim$10,000---35,000\,km in AIA~171\,\AA, 
$\sim$7000---25,000\,km in \iris\ UV; 
width: $\sim$1000---7000\,km, in both AIA~171\,\AA\ and \iris\ UV\@.  {\it Lifetimes}: tens of seconds to $\sim$5\,min.  
{\it Energies}:  Not provided in listed sources.  {\it Locations}: Magnetic network.  
{\it SEIM Objectives}: There is currently some evidence that jetlets are small-scale coronal jets, but this is not yet 
established.  SEIM, in conjunction with other instruments, will continue to investigate whether they 
have properties analogous to coronal jets, UV network jets, and possibly other features.

\noindent
{\bf UV Network Jets}:  {\it Chief references for following parameters}: \citet{tian.et14}.  
{\it Wavelength}: UV\@.   {\it Appearance}: Spire that grows to be long and
narrow, similar to coronal jets but smaller in size.  {\it Size}: length: $\sim$4000---10,000\,km; 
width: $\sim$300\,km.  {\it Lifetimes}: 20---80\,s.  {\it Energies}:  Energy of individual event not provide
in \citet{tian.et14}, but likely $\sim10^{24}$---$10^{26}$\,erg, given that a typical one is slightly larger than 
a typical spicule.  {\it Locations}: Magnetic network.  
{\it SEIM Objectives}: It is unknown whether these features are small jetlets, relatively large 
spicules, or something different from both.  SEIM will not observe in UV, but in conjunction with other 
instruments (IRIS, or a similar instrument) SEIM will investigate whether these features 
have properties analogous to coronal jets and/or jetlets, including brightenings detectable in EUV
at their bases.

\noindent
{\bf Spicules}:  {\it Chief references for following parameters}: \citet{beckers68,sterling00,depontieu.et07a}; \citeauthor{hinode.et19}~(\citeyear{hinode.et19}), 
T. Pereira subsection; \citet{samanta.et19}.  {\it Wavelength}: Traditionally seen in chromospheric lines.  Similar features 
are seen at UV wavelengths
too, at least some of which are components of the chromospheric features.   {\it Appearance}: chromospheric-material 
spire that grows to be long and narrow.  {\it Size}: maximum length: $\ltsim$5000\,km; 
width, $\sim$(1---few)\,$\times 10^2$\,km.  {\it Lifetimes}: tens of seconds to a few minutes. {\it Energies}: $\ltsim 10^{25}$\,erg.
 {\it Locations}: Magnetic network.  {\it SEIM Objectives}: It has been suggested that
some spicules might be scaled-down versions of coronal jets \citep{sterling.et16a,sterling.et20b}.  If this is
the case, then some such spicules might be expected to have brightenings at their bases, some of which
might be visible in EUV.\@  SEIM will look for such brightenings, and compare with similar observations
at the bases of coronal jets, jetlets, and similar features.

\noindent
{\bf Penumbral jets}:  {\it Chief references for following parameters}:   \citet{katsukawa.et07,tiwari.et16,tiwari.et18}.  
{\it Wavelength}: Until now, most 
observations are from \hinode/Solar Optical 
Telescope (SOT) \caii\,H images and \iris\ UV data.  {\it Appearance}: transient, long and very narrow, faint
features seen in sunspot penumbra at high spatial resolution.   {\it Size}: maximum length: 
typically 1000---4000\,km, but some extend up to 10,000\,km; width, $\ltsim$400---600\,km.  
{\it Lifetimes}: $<$1\,min.  {\it Locations}: Sunspot penumbrae.  {\it SEIM Objectives}: \citet{tiwari.et18} suggest
that penumbral jets might form from magnetic processes similar to those suspected of causing coronal 
jets.  SEIM will likely not observe penumbral jets directly due to their suspected low temperature (i.e., little or
no emission in coronal EUV is expected from them).  But a better characterization of properties of coronal jets
and similar features with SEIM, in conjunction with other instruments, will provide fresh data with which 
to compare penumbral filaments, to help ascertain how closely their production mechanism might
resemble that of coronal jets.

\noindent
{\bf Campfires}:  {\it Chief references for following parameters}:   \citet{panesar.et21,berghmans.et21}
{\it Wavelength}: Until now, EUV images from SO/EUI~174\,\AA, and from AIA; and magentograms from 
HMI\@.  {\it Appearance}:  small-scale, 
short-lived coronal brightenings, that can appear as loop-like, dot-like, jet-like, or complex structures.   
{\it Size}: length: $\sim$5000\,km; width: $\sim$1500\,km.  {\it Lifetimes}: $\sim$1---60\,min; 
most $<$10\,min.  {\it Locations}: So far only studied in quiet Sun and polar regions, but observations 
have so far been limited.  They occur at the the edge of magnetic network lanes.  {\it SEIM Objectives}:  \citet{panesar.et21} 
find campfires to have similarities to coronal jets or pre-coronal-jet minifilament regions, in that they occur on 
magnetic neutral 
lines and most contain transient cool-plasma structures.  SEIM observations, in conjunction 
with other instruments, will clarify whether some or all campfires are powered in the manner in which 
coronal jets and pre-coronal-jet X-ray-bright locations are powered.  More generally, some campfires appear similar
to small-scale brightenings called {\it X-ray bright points} (XBPs) and/or X-ray bright-point flares, that have been 
observed since the 1970s \citep{golub.et74}, with the campfires perhaps being smaller-scale and less-energetic 
versions of XBP brightenings; and some of them may be small-scale coronal jets.  SEIM, along 
with complementary instruments, will be able to observe both XBPs and campfires, and of course coronal jets,
providing information to determine whether these features operate by the same or different physical mechanisms.

\newpage

\section*{Figures}


\clearpage

\end{document}